\documentclass[11pt]{article}
\usepackage{blois,epsfig}

\def\be{\begin{equation}}
\def\ee{\end{equation}}
\def\bea{\begin{eqnarray}}
\def\eea{\end{eqnarray}}

\begin{document}
\title{COSMOLOGY WITH PRIMORDIAL BLACK HOLES}

\author{Ga\"elle Boudoul \& Aur\'elien Barrau}

\address{Laboratory for Subatomic Physics and Cosmology (CNRS/UJF)\\
53 Ave des
Martyrs - F-38026 Grenoble Cedex - France}

\maketitle\abstracts{
Some aspects of Cosmology with primordial black holes are briefly reviewed}

\section{Density upper limits}

Small black holes could have formed in the early Universe if the density
contrast was high enough (typically $\delta > 0.3 - 0.7$ , depending on models). 
Since it was discovered by Hawking \cite{Hawking5} that they should evaporate with
a black-body like spectrum of temperature $T=\hbar c^3/(8\pi k G M)$, the emitted
cosmic rays have been considered as the natural way, if any, to
detect them. Those with initial masses smaller than $M_*\approx 5\times
10^{14}~{\rm g}$ should have finished their evaporation by now whereas those with
masses greater than a few times $M_*$ do emit nothing but extremely low energy
massless fields. 

As was shown by MacGibbon and Webber \cite{MacGibbon1}, when the
black hole temperature is greater than the quantum chromodynamics confinement 
scale $\Lambda_{QCD}$, quark and gluon jets are emitted instead of composite 
hadrons. This should be taken into account when computing the cosmic-ray fluxes
expected from their evaporation. Among all the emitted particles, two species are
especially interesting : gamma-rays around 100 MeV because the Universe is very
transparent to those wavelengths and because the flux from PBHs becomes softer 
($\propto E^{-3}$ instead of $\propto E^{-1}$)
above this
energy, and antiprotons around 0.1-1 GeV because the natural background due to
spallation of protons and helium nuclei on the interstellar medium is very
small and fairly well known.\\

Computing the contributions both from the direct electromagnetic emission
and from the major component resulting from the decay of neutral pions, the
gamma-ray spectrum from a given distribution of PBHs can be compared with measurements from the EGRET 
detector onboard the CGRO satellite. This translates (in units of critical density) 
into \cite{Carr1998} : $\Omega_{PBH}(M_*) < 10^{-8}$, improving substantially previous 
estimates \cite{MacGibbon1991}. The expected gamma-ray background has also
recently been taken into account to decrease this upper limit by a factor of
three. Computing the contribution from unresolved blazars and the emission
from normal galaxies, Pavlidou \& Fields \cite{pav} have estimated the
minimum amount of extragalactic gamma-rays which should be expected. The first
one was computed using the Stecker-Salamon model and the second one is assumed
to be proportional to the massive star formation rate (which is itself
proportional to the supernovae explosion rate) as it is
due to cosmic-ray interactions with diffuse gas. 
With a very conservative errors
treatment this leads to \cite{barrauicrc}: $\Omega_{PBH}(M_*)<3.3\times 10^{-9}$.\\

On the other hand, galactic antiprotons allowed new complementary upper limits as
no excess from the expected background is seen in data \cite{maki}. 
The main improvement of the
last years is, first, the release of a set of high-quality measurements from the BESS, 
CAPRICE and AMS experiments. The other important point is a dramatic improvement in
the galactic cosmic-rays propagation model \cite{Webber}. A most promising approach is
based on a two-zone description with six free parameters : $K_0$,
$\delta$ (describing the diffusion coefficient $K(E)=K_0 \beta 
R^{\delta}$), the diffusive halo half height L, the convective velocity $V_c$ 
and the Alfv\'en velocity $V_a$ \cite{Maurin2002}. The latters have been varied within a 
given range determined by an exhaustive and systematic  study of cosmic
ray nuclei data \cite{Maurin2001}, giving a very high confidence in the resulting
limit : $\Omega_{PBH} < 4\times 10^{-9}$ for an average halo size \cite{Barrau2002}.
This approach paid a great attention to many possible sources of uncertainties, 
some being shown to be very small ({\it e.g.} the flatness of the dark matter
halo and its core radius, the spectral index of the diffusion coefficient, the
number of sources located outside the magnetic halo), other being potentially
dramatic ({\it e.g.} a possible "photosphere" around PBHs \cite{Heckler} which could lead to
substantial interactions between partons, a cutoff in the mass spectrum if the
inflation reheating temperature was to low). Furthermore, this study can be improved
by taking advantage of the different ways in which solar modulation changes the primary
and secondary fluxes \cite{Mitsui}. \\

Improvements in the forthcoming years can be expected in several directions.
Gamma-rays could lead to better upper limits with new data from the GLAST satellite
\cite{GLAST} to be launched in 2007. The background due to known gamma-ray sources could also be
taken into account in the analysis. On the other hand, antiprotons should be much better
measured by the AMS experiment \cite{Barrau2001}, to be operated on the
International Space
Station as of 2006. Finally, antideuterons could open a new window for detection. The
probability that an antiproton and an antineutron emitted by a given PBH merge 
into and antideuteron within a jet can be estimated through the popular nuclear physics 
coalescence model. The main interest of this approach is that the spallation
background is extremely low below a few GeV for kinematical reasons (the threshold
is much higher in this reaction (17 $m_p$) than for antiprotons), making the signal-to-noise ratio
potentially very high \cite{Barrau2002b} with a possible improvement in sensitivity
by one order of magnitude.\\

An interesting alternative could be to look for the expected extragalactic neutrino
background from PBHs evaporation \cite{Bugaev}. This is an original and promising 
idea but it suffers from a great sensitivity to the assumed mass spectrum.\\

A very different approach would be to look the gravitational waves emitted by coalescing
PBH binaries \cite{Nakamura}. This could be the only possible way to detect very heavy PBHs
(above a fraction of a solar mass for the current LIGO/VIRGO experiments) 
which do not emit any particle by the Hawking mechanism. If most of the halo dark matter was
made of, say 0.5$M_{\odot}$ PBHs (as suggested in some articles by the MACHO
collaborations that are nowadays disfavoured by the EROS results), there should be about $5\times 10^8$ binaries out to 50 
kpc away from the Sun, leading to a measurable amount of coalescence by the next generation 
of interferometers.

\section{PBH formation in cosmological models}

Small Black holes are a unique tool to probe the very small cosmological scales, 
far beyond the microwave background (CMB) or large scale structure (LSS) measurements
(see \cite{khlobook} for a review). 
In the standard mechanism, they should form with masses close to the horizon mass at a
given time:
$$M_{PBH}(t)=\gamma^{3/2}M_{Hi}\left( \frac{t}{t_i} \right) = \gamma^{3\gamma /
(1+3\gamma)} M^{(1+\gamma)/(1+3\gamma)}M_{Hi}^{2\gamma/(1+3\gamma)}$$

where $M_H$ is the horizon mass, $\gamma$ is the squared sound velocity,
the subscript 'i' represents the quantity at $t_i$, the
time when the density fluctuation develops, and $M\propto M_H^{3/2}$ is the mass
contained in the overdense region with comoving wavenumber $k$ at $t_i$. In principle,
it means that PBHs can be formed with an extremely wide mass range, from the Planck mass
($\approx 10^{-5}$ g) up to millions of solar masses. Assuming that the fluctuations have a
Gaussian distribution and are spherically symmetric, the fraction of regions of mass M
which goes into black holes can (in most cases) be written as :

$$\beta(M)\approx\delta(M)exp\left( \frac{-\gamma^2}{2\delta(M)^2}\right)$$

where $\delta(M)$ is the RMS amplitude of the horizon scale fluctuation. The
fluctuations required to make the PBHs may either be primordial or they may arise
spontaneously at some epoch but the most natural source is clearly inflation
\cite{khlo}. Using the direct (non)detection of cosmic-rays from PBHs, 
the entropy per baryon, the possible destruction of Helium nuclei and some subtle
nuclear process, several limits on $\beta(M)$ have been recently re-estimated
\cite{Carr1998} \cite{Liddle}. The most stringent one comes from gamma-rays for
$M\approx 5\times 10^{14}$ g : $\beta < 10^{-28}$.\\

Such limits can be converted into relevant constraints on the parameters of the
primordial Universe. In particular, a "too" blue power spectrum ($P(k)\propto k^n$ with $n>1$)
would lead to an overproduction of PBHs. Taking advantage of the normalization given 
by COBE measurements of the fluctuations at large scales, a blue power spectrum can be
assumed (as favoured in some hybrid inflationary scenarii for example) and compared with
the gamma-ray background as observed with the EGRET detector \cite{Kim1999}. The
resulting upper limit is $n<1.25$, clearly better than COBE, not as good as recent CMB
measurements \cite{benoit} \cite{map} but extremely important anyway because directly linked with
very small scales. Some interesting developments were also achieved by pointing out a
systematic overestimation of the mass variance due to an incorrect transfer function
\cite{Po1} \cite{Po2}, leading to a slightly less constraining value : $n<1.32$. This
approach also allowed to relax the usual scale-free hypothesis and to consider a step in
the power spectrum, as expected in Broken Scale Invariance (BSI) models : too much power
on small scales (a ratio greater than $\approx 8\times 10^4$) would violate
observations. The highly probable existence of a cosmological constant
($\Omega_{\Lambda}\approx 0.7$) should also be considered to estimate correctly the mass
variance at formation time which should be decreased by about 15\% \cite{Po3}. More
importantly, this approach showed that in the usual inflation picture, no cosmologically
relevant PBH dark matter can be expected unless a well localized bump is assumed in the
mass variance (either as an {\it ad hoc} hypothesis, either as a result of a jump in the
power spectrum, or as a result of a jump in the first derivative of the inflaton
potential) \cite{Blais}. Nevertheless, if the reheating temperature limit due to entropy
overproduction by gravitinos decay is considered \cite{khlolinde}, a large window remains opened for dark
matter \cite{Barrau2003}. The main drawback of PBHs as a CDM candidate is the high level of fine
tuning required to account for the very important dependence of the $\beta$ fraction as a
function of the mass variance.

An important constraint can also be obtained thanks to a possible increase of the PBH 
production rate during
the preheating phase if the curvature perturbations on small scales are sufficiently large. 
In the two-field preheating model with quadratic potential, many values of the 
inflaton mass and coupling are clearly excluded by this approach \cite{Green} :
the minimum preheating duration for which PBHs are overproduced, $m\Delta t$, is of order 60
for an inflaton mass around $10^{-6}$ and a coupling $g\approx 10^{-4}$.\\

Interestingly, the idea \cite{Choptuik} that a new type of PBHs, with masses scaling as $M_{PBH}
=kM_H(\delta_H-\delta_H^c)^{\gamma_s}$, could exist as a result of near-critical collapse,
was revived in the framework of double inflationary models. Taking into account the
formation during an extended time period, an extremely wide range of mass spectra can be obtained
\cite{Kim}. Such near-critical phenomena can be used to derive very important cosmological
constraints on models with a characteristic scale, through gamma-rays \cite{Yoko} or through
galactic cosmic-rays \cite{Barrau2002c}. The resulting $\beta_{max}(M)$ is slightly lower
and wider around $M_*\approx 5\times 10^{14}$ g than estimated through "standard"
PBHs.\\

Alternatively, it was also pointed out that some new inflation models containing one inflaton
scalar field, in which new inflation follows chaotic inflation, could produce a substantial
amount of PBHs through the large density fluctuations generated in the beginning of the second
phase \cite{Yoko2}. Another important means would be to take into account the cosmic QCD
transition : the PBH formation is facilitated due to a significant decrease in pressure
forces \cite {Jedamzik}. For generic initial density perturbation spectra, this
implies that essentially all PBHs may form with masses close to the QCD-horizon scale,
$M_H^{QCD}\approx 1M_{\odot}$.\\

Finally, PBHs could help to solve a puzzling astrophysical problem : the origin of supermassive
black holes, as observed, {\it e.g.}, in the center of active galactic nuclei \cite{Bean}.
In such models, very small black holes attain super massive sizes through the accretion of a
cosmological quintessential scalar field which is wholly consistent with current observational constraints.
Such a model can generate the correct comoving number density and mass distribution and, if
proven to be true, could also constraint the required tilt in the power spectrum. 

Another interesting way to produce massive primordial black holes would be
the collapse of sufficiently large closed domain wall produced during a
second order phase transition in the vacuum state of a scalar field
\cite{khlo2}.

\section{PBH cosmology and fundamental physics}

If the evaporation process was detected, it would mean that 
the Hubble mass at the reheating time was
small enough not to induce a cutoff in the PBH mass spectrum which
would make light black holes exponentially diluted. As
some hope for future detection is still
allowed thanks to antideuterons, it could be possible to give a lower limit on the reheating
temperature \cite{barrauponthieu}.  When compared with the upper limit
coming from Big-Bang nucleosynthesis \cite{kawa}, this translates into
a lower limit on the gravitino mass. It is important to point out that such 
possible constraints
on the gravitino mass can be converted into constraints on more
fundamental parameters, making them very valuable in the search for
the allowed parameter space in {\it grand unified} models. As an
example, in models leading naturally to mass scales in the
$10^2$-$10^3$~GeV range through a specific dilaton vacuum
configuration in supergravity, the gravitino mass can be related with
the GUT parameters \cite{tkach}: $$m_{3/2}=\left(
\frac{5\pi^{\frac{1}{2}}\lambda}{2^{\frac{3}{2}}}\right)^{\sqrt{3}}
(\alpha_{GUT})\left(\frac{M_{GUT}}{M_{Pl}}\right)^{3\sqrt{3}}M_{Pl}.$$
With $M_{GUT}\sim 10^{16}$~GeV and a gauge coupling $\alpha_{GUT}\sim
1/26$, antideuterons could probe an interesting mass range
\cite{barrauponthieu}.\\

Another important challenge of modern physics is to build links between
the theoretical superstring/M-theory paradigm on the one side and the four-dimensional
standard particle and cosmological model on the other side. PBHs can play an important role
in this game. One of the promising way to
achieve a semiclassical gravitational theory is to study the action expansion in scalar
curvature. At the second order level, according to the pertubartional approach of string
theory, the most natural natural choice is the 4D curvature invariant Gauss-Bonnet approach:
$$
S  =  \int d^4 x \sqrt{-g} \Biggl[ - R + 2 \partial_\mu \phi
      \partial^\mu \phi
   +  \lambda e^{-2\phi} S_{GB} + \ldots \Biggr],
$$
where $\lambda$ is the string coupling constant, $R$ is the Ricci scalar, $\phi$
is the dilatonic field and  $S_{GB} = R_{ijkl}R^{ijkl} -
4 R_{ij}R^{ij} + R^2$. This
generalisation of Einstein Lagrangian leads to the very important result that there is a
minimal mass $M_{min}$ for such black holes \cite{c07}.
Solving  the  equations  at first perturbation order with the curvature
gauge metric, it leads to a  minimal radius :
\begin{eqnarray}
r_h^{inf} = \sqrt{\lambda} \ \sqrt{4 \sqrt{6}}  \phi_h (\phi_\infty),
\end{eqnarray}
where $\phi_h (\phi_\infty)$ is the dilatonic value
at $r_h$. The crucial point is that this result remains true when higher order
corrections or time perturbations are taken into account. The resulting value should
be around $2 M_{Pl}$. It is even increased to $10 M_{Pl}$ if moduli
fields are considered, making the conclusion very robust and conservative. In this
approach, the Hawking evaporation law must be drastically modified and an asymptotically
"quiet" state is reached instead of the classical quadratic divergence \cite{Alex}. Although
not directly observable because dominated by the astrophysical background, the integrated
spectrum features very specific characteristics. \\

A very exciting possibility would be to consider particle colliders as (P)BH factories. If the
fundamental Planck scale is of order a TeV, as the case in some extradimension scenarii,
the LHC would produce one black hole per second \cite{Dim} \cite{Gid}. In a brane-world in
which the Standard Model matter and gauge degrees of freedom reside on a 3-brane within a
flat compact volume $V_{D-4}$, the relation between the four-dimensional and the
D-dimensional Newton's constants is simply $G_N=G_D/V_{D-4}$ and
$M_{Pl}^{D-2}=(2\pi)^{D-4}/(4\pi G_D)$. In the - not so obvious for colliders - low angular 
momentum limit, the hole radius, Hawking temperature and entropy can be written as :
$$R_H=\left( \frac{4(2\pi)^{D-4}M}{(D-2)\Omega_{D-2}M_{Pl}^{D-2}} \right) ^{1/(D-3)}~,~
T_H=\frac{D-4}{4\pi R_H}~,~S=\frac{R_H^{D-2}\Omega_{D-2}}{4G_D}$$
where 
$$\Omega_{D-2}=\frac{2\pi^{(D-2)/2}}{\Gamma(\frac{D-1}{2})}$$
is the area of a unit $D-2$ sphere. Together with the black hole production cross section
computation (as a sum over all possible parton pairings with $\sqrt{s}>M_{Pl}$), those
results lead to very interesting observable predictions which - in spite of the "sad
news" that
microphysics would be screened - should allow to determine the number of large new
dimensions, the scale of quantum gravity and the higher dimensional Hawking law. A direct
consequence of those ideas is to look also for black hole production through cosmic-rays interactions 
in the Earth atmosphere. Limiting ourselves with neutrinos - to avoid diffractive phenomena
- it seems that the AUGER, EUSO and OWL experiments should be sensitive to such effects
\cite{Feng} \cite{Dutta}.

\end{document}